\documentclass[12pt]{elsart}

\usepackage{epsfig,amsfonts,amssymb}

\newcommand{\dd}{\mathrm{d}}
\def\bbox#1{{ \mbox{\boldmath $#1$}} }

\begin{document}
\runauthor{Carmona and Cort\'es}
\begin{frontmatter}
\title{Testing Lorentz invariance violations in the
tritium beta-decay anomaly}

\author[Pisa]{J.M. Carmona\thanksref{mail1}},
\author[Zgza]{J.L. Cort\'es\thanksref{mail2}}
\address[Pisa]{Dipartimento di Fisica dell'Universit\`a and INFN,
I-56127 Pisa, Italy}
\address[Zgza]{Departamento de F\'{\i}sica Te\'orica, Universidad de Zaragoza,
50009 Zaragoza, Spain}
\thanks[mail1]{E-mail: carmona@df.unipi.it}
\thanks[mail2]{E-mail: cortes@posta.unizar.es}

\begin{abstract}
We consider a Lorentz non-invariant dispersion relation for the neutrino,
which would produce unexpected effects with neutrinos of few eV, exactly 
where the tritium beta-decay anomaly is found. We use this anomaly to put
bounds on the violation of Lorentz invariance. We discuss other 
consequences of this non-invariant dispersion relation in neutrino experiments
and high-energy cosmic-ray physics.
\end{abstract}



\end{frontmatter}

\section{Introduction}

Recent research on the determination of neutrino mass by studying the 
low-energy beta decay spectrum of tritium has produced a best fit value
for $m^2_\nu$ which is significantly negative~\cite{lobashev}. 
This unphysical value is caused by an anomalous excess of electron 
events at the end of the spectrum, at about 20 eV below the end point.
The origin of this anomaly is not known. It seems clear that there is
some systematic effect not taken into account, most probably of experimental 
nature. 
However, in this letter we want to point that an \textsl{apparent}
excess of electron events near the end of the spectrum is compatible with
a certain deviation of the relativistic dispersion relation for the 
neutrino. In this way, the tritium experimental results could be used to put
bounds on the parameters characterizing this violation of Lorentz symmetry.

The idea that Lorentz invariance is not an exact symmetry, but an approximate
one which works extremely well in our low-energy experiments, is not new.
In the third section of this letter we will review the main proposals for
modifications of the energy-momentum
relativistic dispersion relation. There we will argue that a plausible
dispersion relation for the neutrino, and only for this particle, is
the following one:
\begin{equation}
E_\nu^2=\bbox{p_\nu}^2+m_\nu^2+2\lambda|\bbox{p_\nu}|,
\label{disprel}
\end{equation}     
where $|\bbox{p}|$ means the module of the three-momentum $\bbox{p}$, $m_\nu$
is the neutrino mass, and $\lambda$ is some mass scale, to be determined
afterwards. We will consider $\lambda>0$ in order to have a positive 
contribution to the energy squared.

\section{Kurie plot}

Let us see how the usual Kurie plot of the tritium beta decay is modified
by the dispersion relation Eq.~(\ref{disprel}). The phase-space factor is
\begin{equation}
{\dd}\Pi=p_e^2 {\dd} p_e p_\nu^2 {\dd} p_\nu \, \delta({Q-E-E_\nu}) \ ,
\end{equation}
where $p_e$ is the momentum of the electron, $E$ is the kinetic energy of
the electron ($E_e=m_e+E$) and $Q$ is the energy available to distribute
between the neutrino energy and the kinetic energy of the electron. $Q$ is
given by
\begin{equation}
M-M'=Q+m_e \ ,
\end{equation}
$M$ and $M'$ being the masses of the initial and final nuclei. We are 
neglecting here the kinetic energy of the nucleus, which is typically of order
$10^{-1}$ eV. 
The Kurie plot is proportional to the function $K(E)$ given by
\begin{equation}
K(E)=\left[ \int \delta({Q-E-E_\nu}) \, p_\nu^2 {\dd} p_\nu\right]^{1/2} \ .
\end{equation}
For the usual relativistic dispersion relations, 
$E_e^2=\bbox{p_e}^2+m_e^2$ and 
$E_\nu^2=\bbox{p_\nu}^2+m_\nu^2$, one gets
\begin{equation}
K(E)=\left[(Q-E)\sqrt{(Q-E)^2-m_\nu^2}\right]^{1/2} \ ,
\label{standardK}
\end{equation}
which is a straight line, $(Q-E)$, in the case $m_\nu=0$.

If one takes the new dispersion relation for the neutrino, Eq.~(\ref{disprel}),
then the function $K(E)$ becomes (from now on, we will use the notation 
$m\equiv m_\nu$)
\begin{eqnarray}
K(E)&=&(Q-E)^{1/2}\left\{\left[(Q-E)^2+\lambda^2-m^2\right]^{1/4}\right. 
\nonumber \\
&& -
\left.\frac{\lambda}{\left[(Q-E)^2+\lambda^2-m^2\right]^{1/4}}\right\} \ .
\end{eqnarray}
It is easy to check that the end point of this curve is $E=Q-m$, just
as in the standard case Eq.~(\ref{standardK}). On the other hand, for large
$Q-E$ values, the curve is very well approximated by a straight line. 
Considering a point $E_l$ so that $Q-E_l \gg m,\lambda$, we introduce the
linear approximation
\begin{equation}
\bar K_{E_l}(E) = K(E_l)+\left.\frac{\partial K(E)}{\partial E}\right|_{E=E_l}
(E-E_l) \ ,
\end{equation}
i.e., it is the tangent to $K(E)$ at the point $E=E_l$.
Expanding $\bar K_{E_l}(E)$ in powers of $\lambda/(Q-E_l)$, 
$m/(Q-E_l)$, we get
\begin{equation}
\bar K_{E_l}(E) = (Q-E)-\lambda + \frac{\lambda^2-m^2}{4(Q-E_l)} 
+ \frac{\lambda^2-m^2}{4(Q-E_l)^2}(E-E_l) + \ldots
\end{equation}
so that, instead of having a straight line of slope $-1$ ending at
$E=Q$, which is the standard case with $m=\lambda=0$, we obtain a
straight line of slope
\begin{equation}
-1+\frac{1}{4}\frac{\lambda^2-m^2}{(Q-E_l)^2} + \ldots
\end{equation}
which ends at the point
\begin{equation}
E\equiv E_0 = Q-\lambda +\frac{1}{2}\frac{\lambda^2-m^2}{Q-E_l} + \ldots
\end{equation}
On the other hand, one can calculate the exact value of the slope
at the end point:
\begin{equation}
\left.\frac{\partial K(E)}{\partial E}\right|_{E=Q-m} = 
-\left(\frac{m}{\lambda}\right)^{3/2} \ .
\end{equation}

\begin{figure}[tb]
\centerline{\epsfig{figure=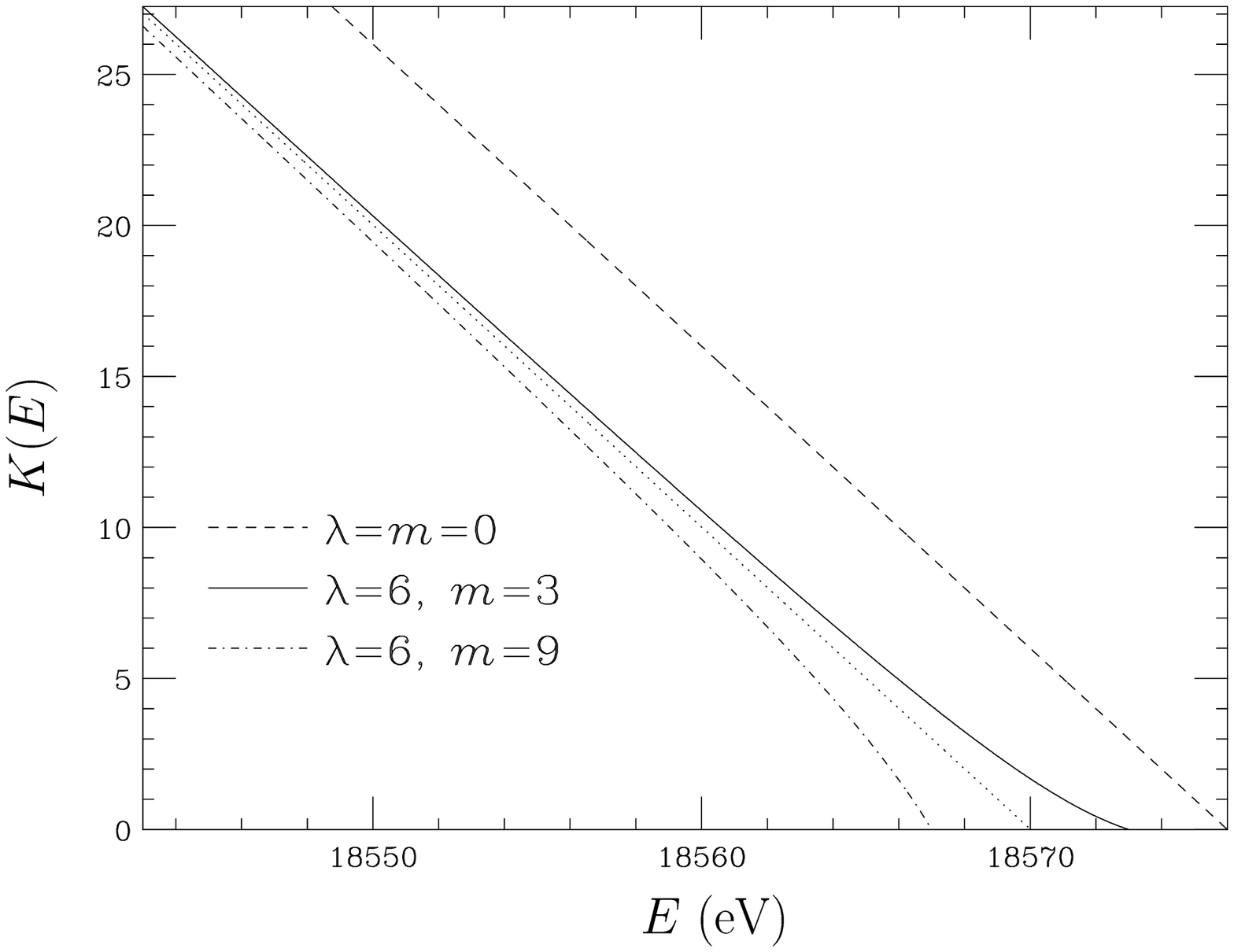,angle=90,width=11cm}}
\caption{The function $K(E)$ for the two cases $\lambda>m$ and $\lambda<m$,
together with their linear approximation (dotted line). The straight line
$Q-E$, corresponding to $\lambda=m=0$ is also shown. Units in this 
figure are eV.}
\label{fig:curvas}
\end{figure}

Two cases are clearly distinguished: $\lambda>m$ and $\lambda<m$.
These are shown in Fig.~\ref{fig:curvas}. We see that near the end of 
the spectrum, the curve $K(E)$ is above the linear approximation
$\bar K_{E_l}(E)$ when $\lambda>m$, which corresponds to an
\textsl{apparent} excess of electrons at high energies. Indeed it is 
only apparent, because the curve lies always below the corresponding
curve of a relativistic dispersion relation for a massless neutrino,
which is also indicated in the figure. In the $\lambda<m$ case, we get the
oposite situation: the effect due to the neutrino mass dominates over the
$\lambda$ term (responsible for the ``apparent excess'') and we get a
reduction on the number of electrons at high energies.

The tritium anomaly consists in an excess of electron events at high
energies, where ``excess'' here means that the data stay above the
straight line which is the linear approximation to the curve at low
energies. A possible measure of the anomaly could be given by using
the following quantity
\begin{equation}
R_\Delta=4\frac{\int_{E_0-\Delta/2}^{E_0} {\dd}E \,K(E)}
{\int_{E_0-\Delta}^{E_0} {\dd}E \,K(E)} \ ,
\end{equation}
so that $R_\Delta>1$ indicates an apparent excess of electrons at
high energies. This quantity could be measured experimentally, and
a comparison with the theoretical prediction would put bounds on the
values of the parameters $\lambda,m$. We will show now the orders of
magnitude of these bounds in a hypothetical but realistic example.

\begin{figure}[tb]
\centerline{\epsfig{figure=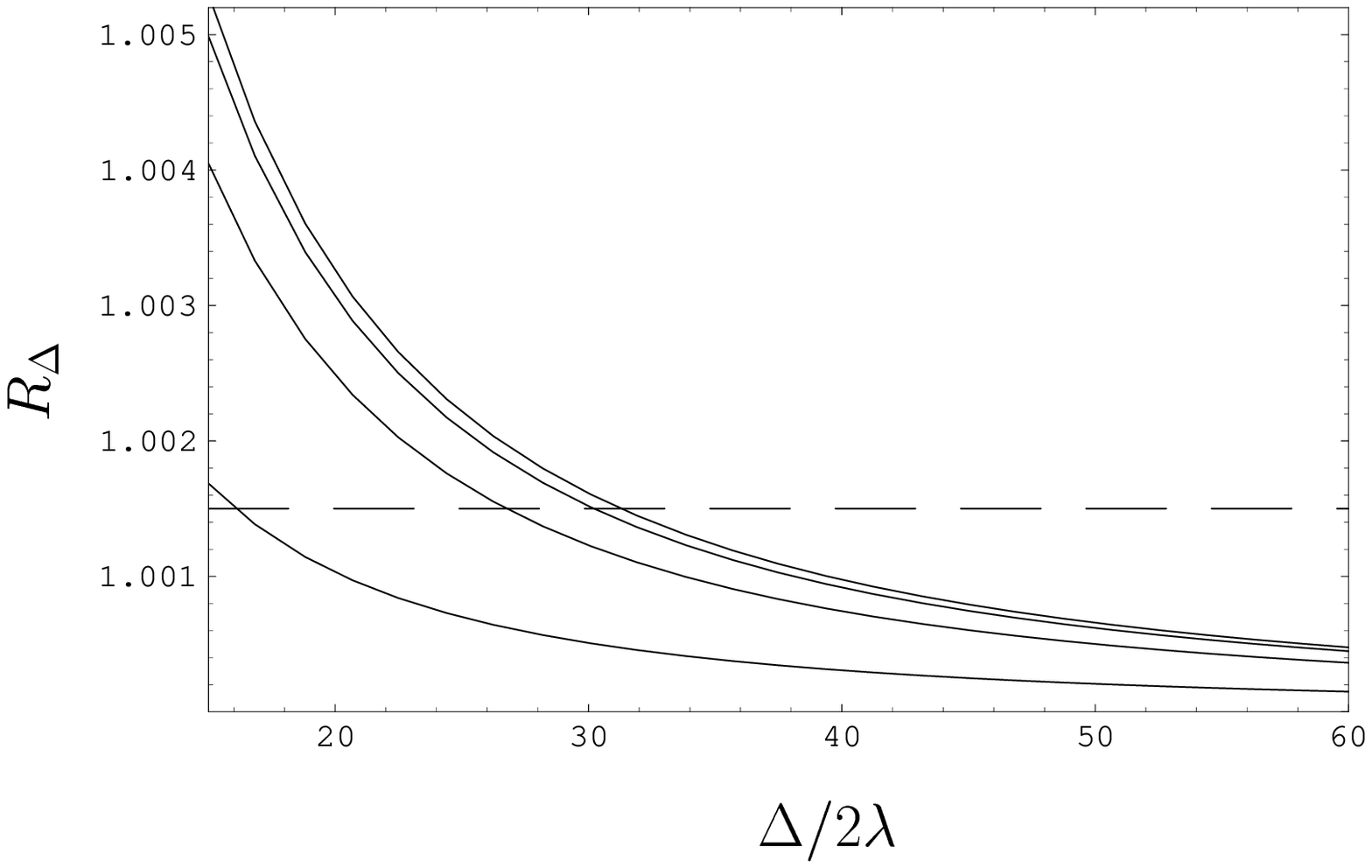,angle=90,width=12cm}}
\caption{$R_\Delta$ as a function of $\Delta/2\lambda$ for three 
different values of $\alpha$: starting from the lower curve, 
$\alpha=0.3$ ($\lambda=1.195 m$), 
$\alpha=0.75$ ($\lambda=2m$), $\alpha=0.9375$ ($\lambda=4m$)
and $\alpha=1$ ($m/\lambda=0$). The dashed line $R_\Delta=1.0015$ is
a hypothetical experimental bound on $R_\Delta$ (see text).}
\label{fig:R}
\end{figure}

Introducing the variable $x=(Q-E)/\lambda$, 
and taking $E_0\approx Q-\lambda$,
$R_\Delta$ is rewritten as follows:
\begin{equation}
R_\Delta=4\frac{\int_1^{1+\frac{\Delta}{2\lambda}} {\dd}x \,
\sqrt{x} \left[(x^2+\alpha)^{1/4}-(x^2+\alpha)^{-1/4}\right]}
{\int_1^{1+\frac{\Delta}{\lambda}} {\dd}x \,
\sqrt{x} \left[(x^2+\alpha)^{1/4}-(x^2+\alpha)^{-1/4}\right]} \ ,
\end{equation}
where $\alpha\equiv 1-m^2/\lambda^2$.
Fig.~\ref{fig:R} shows $R_\Delta$ as a function of $\Delta/2\lambda$ for
different values of $\alpha$, that is, for different values of the
quotient $m/\lambda$.

Experimentally, one could put a bound on the excess of electron events,
that is, write $R_\Delta<1+\epsilon_\Delta$ for a certain value of 
$\Delta$, which should be chosen in an appropiate way. $\Delta$ should be
large enough, so that all the anomaly is contained in the region
$E_0-\Delta/2<E<E_0$, but it should not be too large, because in that
case $R_\Delta\to 1$. Let us do an estimate of orders of magnitude.
From Table 1 of Ref.~\cite{belesev}, we see that the fit which gives
$m^2_\nu<0$, signal of the anomaly, is stable and has a good 
$\chi^2/\mathrm{d.o.f.}$ in a region between 200 and 400 eV before the
end point, which is $E_0\sim 18570$ eV. Let us take $\Delta/2=200$ eV.
An approximate value of $R_\Delta$ comes then from Fig.~2 of 
Ref.~\cite{belesev},
which contains the experimental data for the Kurie plot. 
$R_\Delta$ is approximately given by $4(s+t)/(4s+t)$, where $s$ is the
area of a triangle of base and height $\Delta/2$ (the straight line of
the Kurie plot has slope $-1$), and $t$ is the area delimited by the data
above the straight line, spread in a range of energies of around 20 eV, 
and the line itself. A rough estimate of this area gives us a 
conservative value of $\epsilon_\Delta=1.5\cdot 10^{-3}$. For the four
curves of Fig.~\ref{fig:R}, corresponding to four different values
of the quotient $m/\lambda$, we obtain bounds for $\Delta/2\lambda$,
which, recalling that we have taken $\Delta/2=200$ eV, produce bounds
for $\lambda$ which go from $\lambda<12.5$ eV to $\lambda<6.5$ eV for the
extreme curves. Again, this translates into bounds for the neutrino mass,
depending on the value of $\alpha$. For $\lambda=2m$, the bound is
$m<3.8$ eV, and for $\lambda=4m$, we get $m<1.7$ eV. 

We see that the tritium anomaly puts bounds for $(\lambda,m)$ in a 
very interesting range (a few eV). Of course, a detailed analysis of the
experimental data would give finer bounds to this kind of 
Lorentz invariance violation. Even more interesting would be a possible
(future) experimental bound of the type $R_\Delta>1+\epsilon_\Delta$, with
$\epsilon_\Delta>0$ (that is, the confirmation of the anomaly), which would
give a lower bound for $\lambda$, showing the presence of Lorentz 
invariance violation effects in the tritium beta decay. 

\section{Lorentz invariance violations}

Special relativity and Lorentz invariance are at the base of our 
low-energy effective theories. However, it may be possible that these are
low-energy symmetries of a larger theory that do not need to be
Lorentz invariant. Several attempts have been made to question Lorentz
invariance~\cite{kostelecky} and put bounds on possible violations 
(see e.g.~\cite{coleman}).

One way to explore the potentially observable effects of departures 
from exact Lorentz invariance is to consider the consequences in low-energy 
processes of possible extensions of the Lorentz-invariant particle 
energy-momentum relation compatible with translational and rotational
invariance in a ``preferred frame''. A first possibility is given
by the Lorentz-violating class of dispersion relations~\cite{amelino}
\begin{equation}
E^2 = \bbox{p}^2+m^2+\bbox{p}^2 \left(\frac{|\bbox{p}|}{M}\right)^n \ ,
\label{ep1}
\end{equation}
where $M$ is the natural mass scale of the Lorentz non-invariant fundamental
theory. An analysis 
of cosmic ray processes, whose thresholds are drastically
changed by the new dispersion relation, puts severe bounds on the scale of
the Lorentz violation: $M$ has to be several orders of magnitude larger
than the Planck mass scale~\cite{grillo}.

Another possible extension of the energy-momentum relation, coming from the
introduction of a rotationally invariant two-derivative term in the free
Lagrangian, is~\cite{coleman}
\begin{equation}
E^2 = \bbox{p}^2+m^2+\epsilon\bbox{p}^2 \ ,
\label{ep2}
\end{equation}
where $\epsilon$ is a small coefficient which fixes the maximal attainable
velocity of each particle ($v^2 = 1+\epsilon$). Differences among maximal
attainable velocities of different particles lead to abrupt effects when 
the dimensionless ratio $\epsilon\bbox{p}^2/m^2$ is of order unity. The 
precise tests of special relativity give very strong constraints on this 
type of extension~\cite{coleman} ($\epsilon < 10^{-23}$).   
 
A discussion of departures from exact Lorentz invariance in terms of 
modifications of the energy-momentum relation leads to consider a third
possible extension of the dispersion relation with a term linear in
$|\bbox{p}|$~\cite{alfaro},
\begin{equation}
E^2 = \bbox{p}^2+m^2+2\lambda|\bbox{p}| \ .
\label{ep3}
\end{equation}
The additional term dominates over the standard kinetic term ($\bbox{p}^2$)
when $|\bbox{p}|\lesssim 2\lambda$ and then the nonrelativistic kinematics
is drastically changed. Therefore this type
of generalized dispersion relation has to be excluded, except just for one 
case. The neutrino has two characteristic properties: it has a very small 
mass, and it interacts only weakly. As a result of this combination, we have 
not any experimental result on its nonrelativistic physics. Therefore, the 
presence of Lorentz invariance violations affecting the nonrelativistic limit 
cannot be excluded a priori in the neutrino case.

One possible way to incorporate a departure of Lorentz invariance affecting 
only to the neutrinos is to assume that the presence of a linear term in
the dispersion relation is due to a new interaction which acts as a messenger 
of the Lorentz non-invariance at high energies. Once again, this is not a
weird assumption: the introduction of new interactions is a 
general practice in the different attempts to explain the smallness of neutrino
masses~\cite{mohapatra}.

In conclusion, a dispersion relation of the form~(\ref{disprel}) can
be considered for the neutrino at low energies. We still assume
the existence of further Lorentz noninvariant terms, as those contained
in Eqs.~(\ref{ep1}) and~(\ref{ep2}), for the neutrino
as well as for all other particles. In the rest of the 
letter, we turn our attention to other implications of Eq.~(\ref{disprel}).  

\section{Other implications of the new dispersion relation}

\subsection{Neutrino oscillations}

Let us first concentrate on the influence on a characteristic low-energy
phenomenon for the neutrino: flavour oscillations. In the case of
a Lorentz invariance violation independent of flavour, that is,
\begin{equation}
E_i^2 = \bbox{p}^2 + m_i^2 + 2 \lambda |\bbox{p}| \ ,
\end{equation}
one gets the result
\begin{equation}
E_i-E_j=\frac{m_i^2-m_j^2}{2|\bbox{p}|} + \ldots
\end{equation}
so that there is not any footprint of Lorentz noninvariance in
neutrino oscillations. If we admit a possible dependence of 
$\lambda$ on the flavour, we get
\begin{equation}
E_i-E_j = \lambda_i-\lambda_j-
\frac{(\lambda_i^2-\lambda_j^2)-(m_i^2-m_j^2)}{2|\bbox{p}|} + \ldots
\end{equation}
and the oscillation probability becomes
\begin{eqnarray}
P(\nu_i\to\nu_j)&\approx& \sin^2 2\theta \, \sin^2 
\left(L({\mathrm{Km}})\left[(1.27)\frac{(m_i^2-m_j^2)({\mathrm{eV^2}})}{E({\mathrm{GeV}})}\right.\right.\nonumber\\
&&
\left.\left.+(3.54)10^9 (\lambda_i-\lambda_j)({\mathrm{eV}})\right]\right) \ ,
\end{eqnarray}
which means an energy independent probability (which goes against 
the experimental observations in solar neutrino oscillations) unless
\begin{equation}
|\lambda_i-\lambda_j|\leq 10^{-18} ({\mathrm{eV}}) \ ,
\end{equation}
where we have used that $L_\odot=10^8$ Km, and 
$\lambda_i^2-\lambda_j^2\ll m_i^2-m_j^2$. There will be no footprint
of Lorentz invariance violations in the case of atmospheric neutrinos,
since $L_{\mathrm{atm}}\sim 10^{-4} L_\odot$, nor in
accelerator experiments, with much smaller lengths. The conclusion is
that the linear term in $|\bbox{p}|$ in the generalized dispersion
relation~(\ref{disprel}) has to be flavour-independent.

\subsection{Contribution to the energy density of the Universe}

The bounds on neutrino masses from their contribution to the energy
density of the Universe~\cite{pdg98} are not affected by the
presence of Lorentz invariance violations, since they only depend on
the minimum energy of the neutrino, which is still $E_{\mathrm{min}}=m$
(for $\bbox{p}=0$), independently of $\lambda$. 

\subsection{Neutrinos in astrophysics}

The spread of arrival times of the neutrinos from SN1987A, coupled with the 
measured neutrino energies, provides a simple time-of-flight limit on 
$m_{\nu_e}$.

From Eq.~(\ref{disprel}) one has for the neutrino velocity
\begin{equation}
\beta = \frac{\partial E}{\partial |\bbox{p}|} = 
\frac{1}{\sqrt{1 - \frac{1}{|\bbox{p}|^2} \frac{\lambda^2 - m^2}
{(1 + \lambda/|\bbox{p}|)^2}}} \ .
\label{beta}
\end{equation}
Since $|\bbox{p}| \gg \lambda$, the limit~\cite{pdg98} (23 eV)
can be taken as an upper limit for $\sqrt{\lambda^2-m^2}$ which is not
far away from the range of parameters suggested by the tritium beta-decay
anomaly. 
Then, neutrinos from supernovas can be a good place to look for
footprints of Lorentz invariance violations.
In fact, we see another implication from Eq.~(\ref{beta}) if $\lambda>m$: 
neutrinos travel faster than light, so that those carrying more energy will
be the latest to arrive, in contrast with the case of a relativistic
dispersion relation.

\subsection{Consequences in particle and cosmic ray physics}

A striking consequence of the presence of the $\lambda$-term in the
generalized neutrino dispersion relation Eq.~(\ref{disprel}) is the 
kinematic prohibition of reactions involving neutrinos at a certain energy. 
Let us consider ordinary neutron decay: $n \to p + e^- + \bar \nu$.
At large momentum, the $\lambda$-term, which is proportional to 
$|\bbox{p}|$, represents a large contribution to the energy balance, so that
the process might be forbidden for neutrons of sufficiently high momentum.
This turns out to be the case: it is easy to see that for 
$\lambda/|\bbox{p}|\ll 1$ and $|\bbox{p}|\gg m_n, m_p$, 
the total energy of the final state is bounded from below,
\begin{equation}
E_p + E_e + E_{\bar\nu} > |\bbox{p}| + \lambda + 
\frac{(m_p+m_e)^2}{2|\bbox{p}|} \ ,
\end{equation}
which implies that \textsl{neutrons with momentum}
\begin{equation}
|\bbox{p}|>\frac{m_n^2-(m_p+m_e)^2}{2\lambda} = 
\left[\frac{1.47\cdot 10^{15}}{2\lambda {\mathrm{(eV)}}}\right]{\mathrm{(eV)}}
\end{equation}
\textsl{are stable particles.}
A similar conclusion is obtained in the case of pions: the desintegration
$\pi^- \to \mu^- + \bar\nu_\mu$ is forbidden if the pion has a momentum
\begin{equation}
|\bbox{p}|>\frac{m_\pi^2-m_\mu^2}{2\lambda} = 
\left[\frac{8.3\cdot 10^{15}}{2\lambda{\mathrm{(eV)}}}\right]
{\mathrm{(eV)}} \ .
\end{equation}

This means that pions and neutrons with these energies can form part of
cosmic rays, since they are stable particles if the neutrino energy-momentum
relation is given by Eq.~(\ref{disprel}). Therefore, cosmic ray physics might
be drastically affected at high energies, of order 
$E\gtrsim 10^{15}$--$10^{16}$ eV. 
In particular, the presence of stable neutrons and pions in cosmic rays
at such energies might contribute to avoiding the well-known GZK
cutoff~\cite{GZK}. 

\section{Conclusions}

It is surprising that a very simple extension of the Lorentz invariant 
dispersion relation for the neutrino, consistent with all the
constraints from neutrino physics, 
can affect phenomena of such different energy ranges. It could
be behind the tritium beta-decay anomaly, and also lead to
a drastic change in the composition of cosmic rays
beyond $10^{16}$ eV. It seems worthwhile to explore in more detail all
the consequences of this violation of Lorentz invariance and possible
models of the Lorentz non-invariant physics at high energies incorporating
the extended neutrino dispersion relation considered in this letter as a 
low-energy remnant. 

\begin{ack}
We are grateful to J.L. Alonso for collaboration at early stages of this work,
to D.E. Groom for useful comments on the tritium anomaly, to A. Grillo,
R. Aloisio and A. Galante for discussions on observable effects of 
violations of Lorentz invariance, and to J.M. Us\'on.
The work of JMC was supported by EU TMR program
ERBFMRX-CT97-0122 and the work of JLC by CICYT (Spain) 
project AEN-97-1680.
\end{ack}

\end{document}